\documentclass[nofootinbib, reprint, amsmath,amssymb, aps,]{revtex4-1}
\usepackage{graphicx}
\usepackage{dcolumn}
\usepackage{bm}
\usepackage[colorlinks]{hyperref}

\usepackage{graphicx,times}
\usepackage{subfigure}
\usepackage{amsmath}
\usepackage{mathrsfs}
\usepackage{cases}
\usepackage{bm}
\usepackage{longtable}
\usepackage{hyperref}
\usepackage{epstopdf}
\usepackage{amsmath,bm}
\usepackage{amssymb}
\usepackage{natbib}
\usepackage{morefloats}
\usepackage{multirow}
\usepackage{array}
\usepackage{verbatim}
\usepackage[stable]{footmisc}
\usepackage{stackrel}
\usepackage{graphicx}
\usepackage{dcolumn}
\usepackage{bm}

\newcommand{\br}{{\bf r}}
\newcommand{\by}{{\bf y}}
\newcommand{\bx}{{\bf x}}

\newcommand{\bxi}{{\mbox{\boldmath $\xi$}}}

\begin{document}


\title{Magnetic topology in plasmas}

\author{Amir Jafari}
 \email{elenceq@jhu.edu}%
\affiliation{%
Johns Hopkins University, Baltimore, USA\\
}%

\author{Ethan Vishniac}
 \email{ethan.vishniac@aas.org}
\affiliation{%
Johns Hopkins University, Baltimore, USA\\
}%

\date{April 20, 2020}

\begin{abstract}
We study the evolution of turbulent magnetic fields from a topological point of view, invoking commonplace mathematical tools from general topology and dynamical systems theory which connect magnetic field evolution to time reversal invariance, entropy increase and the second law of thermodynamics. We show that in fact magnetic topology is well-defined only in the phase space corresponding to a dynamical system governed by the induction equation. Hence the field's topology and stochasticity can be studied in terms of the corresponding phase space trajectories rather than the field lines in real Euclidean space. In fact, our results suggest that magnetic field lines should not be taken too literally because their existence and uniqueness and more importantly continuity in time require strong mathematical conditions, hardly satisfied in astrophysical systems. As for magnetic topology change, it is shown that the phase space topology is preserved in time for a magnetic field which, besides satisfying few continuity conditions, solves a time reversal invariant induction equation. What breaks the time symmetry in the induction equation is the presence of non-ideal plasma effects at small scales  such as resistivity, which results from random collisions between diffusing electrons and other particles. The small scale, stochastic disturbances produced thereby are super-linearly amplified by Richardson diffusion in the turbulent cascade, which are eventually manifested as large scale reconnection events, somehow similar to stretching quantum fluctuations during inflation to seed large scale cosmological structures. This suggests that reconnection is rooted in the second law of thermodynamics that dictates entropy increase which in turn breaks the time symmetry. 


\end{abstract}

\pacs{Valid PACS appear here}
\maketitle


\section{Introduction}

The sun, studied more than any other star for obvious reasons, has long played an important role in our attempts to understand nature, from helium which was first discovered on the sun to the evolution of magnetic fields in electrically conducting fluids. The internally generated solar magnetic fields, by the magnetic dynamo action working probably near the tachocline, affect the dynamics on the solar surface, e.g., the evolution of sunspots, and also more distant structures including our communication systems and, as a matter of fact, the biosphere. The sun's magnetic signature is also carried over into space by means of a stream of high energy charged particles, known as the solar wind, whose interaction with the magnetosphere leads to phenomena such as aurora (northern lights).  Different magnetic processes observed on the sun, the corona and in the solar wind are also found abundantly in a spectrum of other systems from controlled fusion devices to the accretion disks around massive black holes. One example of such ubiquitous processes is the spontaneous acceleration of fluid particles in regions with strong magnetic shear, a process dubbed magnetic reconnection. A variety of non-ideal plasma effects such as resistivity (e.g., see \cite{Schindleretal1988}; \cite{Biskamp1996}; \cite{Priestetal2007}; \cite{Yamadaetal2010}) as well as different non-linear turbulent effects (e.g., see \cite{LazarianandVishniac1999};  \cite{Eyink2015}; \cite{Jafarietal2018}; \cite{JafariandVishniac2018}) have been proposed as mechanisms driving reconnection. These mechanisms are believed to enforce magnetic field lines to disconnect and reconnect again giving rise to a different field configuration which has a lower energy---a process of energy relaxation. The rapidly reconnecting field lines, as widely believed, can accelerate fluid elements on their way, whose collective effect at larger scales may be observed as spontaneous, eruptive fluid motions. There exists, of course, a large number of competing reconnection models, some more plausible than the others. For instance, the rapid motions caused by reconnection can make the flow turbulent (\cite{LazarianandVishniac1999}; \cite{JV2019}), even if it was initially laminar, which demands taking into account the effects of turbulence, ubiquitous in astrophysical systems (see also \cite{Eyinketal2013}; \cite{Lalescuetal.2015}). Nevertheless, even with such more general considerations, neither of current reconnection models seems to satisfactorily explain observations invoking a coherent formalism. Rather, different models seem to be applicable to different systems. In addition, it is not clear at present what may be the underlying mechanism governing reconnection in terms of the most fundamental laws of physics, not to mention the lack of complete consensus on the definition of reconnection itself. 

Majority, if not all, of magnetic reconnection models are described in some way or another employing the concept of magnetic field lines, put forward long time ago by Faraday\footnote{It seems that for Faraday, magnetic field lines, which he called the lines of magnetic force, had more physical character than what an abstract mathematical object would have. In 1864, Maxwell adopted a slightly different terminology of magnetic force tubes but neither of them, of course, scrutinized the validity or usefulness of such notions for time-dependent fields in magnetized fluids.}. In fact, reconnection is usually described, or even interchangeably referred to, as the magnetic topology change visualized in terms of magnetic field lines. Nonetheless, in the literature of magnetic reconnection, less attention is paid to the mathematical conditions required for magnetic field lines to be well-defined. The same goes for the definition of topology and topology change. Interestingly, in fact, it turns out that the concept of field lines is well-defined only for magnetic fields satisfying a strong mathematical condition (Lipschitz continuity; see \S\ref{section2}), which is violated e.g., in turbulence. In addition, it also turns out that for time dependent fields in dissipative or turbulent media, the magnetic field lines do not continuously evolve in time, i.e., they do not deform continuously in space as time advances and consequently lack a preserved identity. Moreover, as we show in this paper,  magnetic topology is not in fact preserved in time even for well-behaved and smooth fields. The term topology-change too is usually employed in an inaccurate way, paying less attention, if at all, to certain mathematical technicalities involved (see \S\ref{section4}). Finally, in addition to the notion of topology, specially in turbulent reconnection and dynamo models, the concept of weak/strong magnetic stochasticity is also widely used in an almost casual way without providing a precise quantitative measure. These mathematical issues are physically remarkable since they point to fundamental misunderstandings and misrepresentations, which might in fact be responsible for our failure in achieving a fully satisfactory theory for magnetic field reconnection and magnetic field generation in astrophysical objects. Taking for granted the validity of apparently simple mathematical concepts, without actually validating them, can lead us astray.

Reconnection in laminar and turbulent flows is in fact a problem in plasma physics and magnetohydrodynamics which constitutes an intense research field developed primarily by plasma physicists and astrophysicists. In this paper, we approach the more fundamental aspects of this phenomenon as a general problem in theoretical physics to show how mathematical subtleties involved are connected to fundamental laws of physics. We revisit the notions of magnetic field lines, magnetic topology and topology change in particular for stochastic fields in the presence of turbulence. More detailed mathematical results applicable to any physical vector field can be found in \cite{dynamicsJV2019}. In order to quantify the notions of weak and strong magnetic stochasticity, we employ a statistical formalism recently developed in \cite{JV2019}, which was successfully tested in a subsequent work \cite{SecondJVV2019} using an MHD numerical simulation. The magnetic stochasticity level developed in these works was quantitatively related to magnetic diffusion in \cite{JVV2019} which was also accompanied with a numerical test. We start off, in \S\ref{section2}, by a brief, quantitative consideration to show that  a renormalized version of magnetic and velocity fields should be employed in order to study the magnetic field evolution and reconnection in turbulent media. We also show that 
magnetic field lines are not well-defined as continuously deforming curves in space in majority of physical situations e.g., in astrophysical systems. In \S\ref{section3}, we reformulate the problem of magnetic reconnection based on the recent mathematical developments presented in \cite{JV2019}. In \S\ref{section4}, we  show that magnetic topology can be considered in the phase space $(\bf x, B)$ governed by the induction equation instead of the field lines in real Euclidean space. Moreover, the mathematical conditions required to keep the magnetic topology intact are discussed in terms of the induction equation. After establishing a mathematically rigorous notion for magnetic field topology, it is shown in \S\ref{section5} that magnetic topology change is related to the presence of non-ideal plasma effects and the non-linear turbulent effects, which break the time reversal invariance in the momentum and induction equations. This latter statement, in fact, illustrates that reconnection is rooted in the second law of thermodynamics, for the entropy increase in isolated, diffusive systems breaks the time symmetry. We summarize and discuss our results in \S\ref{section6}.

\section{Time Evolution and Field Lines}\label{section2}

From the Maxwell's equations, the governing equation for magnetic field ${\bf B}({\bx}, t)$ is given by

\begin{equation}\label{Faraday}
{\partial {\bf B}\over\partial t}=-\nabla\times{\bf E},
\end{equation}

where ${\bf E}({\bx}, t)$ is the electric field. In a magnetized, and electrically conducting fluid, e.g., a plasma, the momentum equation for electrons dictates the generalized Ohm's law $\bf E+u\times B=P$, hence eq.(\ref{Faraday}) becomes the induction equation:
\begin{equation}\label{Faraday1}
{\partial {\bf{B}}\over \partial t}=\nabla\times ( {\bf{u\times B}-P} ),
\end{equation}
where ${\bf u}={\bf u}({\bx}, t)$ is the velocity field (which is usually a solution of the Navier-Stokes equation although this assumption is not necessary here), and ${\bf P}$ represents non-ideal plasma effects, such as the Hall effect or the resistive electric field (which arises due to electron-ion collisions; ${\bf P}=\eta {\bf J}$ with $\eta$ being resistivity and ${\bf J=\nabla\times B}$ the electric current). In the limit ${\bf P}\rightarrow \bf 0$, e.g., a vanishing resistive electric field in the limit $\eta\rightarrow 0$, one can use the induction equation as $D_t{\bf B}={\bf B.\nabla u-B\nabla. u}+\eta \nabla^2 {\bf B}$ with Lagrangian derivative $D_t\equiv (\partial_t+{\bf u.\nabla}$) and combine it with the continuity equation $D_t\rho+\rho\nabla.{\bf u}=0$ to write $D_t({ {\bf B}/ \rho})=( {{\bf B}/ \rho}).\nabla{\bf u}$ which represents the conventional flux freezing theorem. This result is based on the presumption that MHD equations remain well-behaved in the limit $\eta\rightarrow 0$, hence the integral curves (field lines) of ${\bf B}/\rho$ are advected with the fluid. This approximation, widely used in plasma physics and astrophysics, should be applied with care however---for flux freezing will not hold even as an approximation if the flow becomes turbulent. It turns out, as a matter of fact, that in a fully turbulent fluid, in the limit of vanishing viscosity and resistivity, the velocity and magnetic fields will be in general H{\"o}lder singular\footnote{The real field ${\bf B}(\bx)$ is H{\"o}lder continuous in ${\bx}\in{\mathbb{R}}^n$ if $\|{\bf B}(x) - {\bf B}(y) \| \leq C\| x - y\|^h$ for some $C>0$ and $h>0$. If $h=1$, for any $x, y$, ${\bf B}$ is uniformly Lipschitz continuous. Also ${\bf B}$ is called H{\"o}lder singular if $0<h< 1$. Roughly speaking, a uniformly Lipschitz function $f(x)$ has a bounded derivative, i.e., $|f'(x)|<M$ for some $M>0$. Hence, the derivative of a H{\"o}lder singular function can blow up; $|f'(x)|>\infty$.} which implies ill-defined spatial derivatives (gradients) and hence ill-defined MHD equations\footnote{Experiments and simulations show that viscous energy dissipation rate $\nu|\nabla{\bf u}|^2$ in a turbulent fluid with viscosity $\nu$ and velocity field $\bf u$ does not vanish in the limit $\nu\rightarrow 0$. Similarly, magnetic energy dissipation rate $\eta|\nabla{\bf B}|^2$ in a turbulent fluid with magnetic diffusivity $\eta$ does not approach zero when $\eta\rightarrow 0$. These dissipative anomalies indicate that in the limit $\nu, \eta\rightarrow 0$, the velocity and magnetic divergences blow up, i.e., $|\nabla\bf u|>\infty$ and $|\nabla\bf B|>\infty$. These ultra-violet (UV) divergences, historically first encountered in quantum field theories, imply that turbulent velocity and magnetic fields are not Lipschitz continuous. For details see \cite{Eyink2018}; \cite{JV2019} and references therein.}  (\cite{dynamicsJV2019}; \cite{JV2019}; \cite{Eyinketal2013}).

In order to remove the H{\"o}lder singularities from a given field ${\bf{B}}({\bf{x}}, t)$, defined in a spatial volume $V$, we can coarse-grain or renormalize it at a spatial scale $l>0$ using distributions, e.g., by writing

\begin{equation}\label{coarsegrain1}
{\bf{B}}_l ({\bf{x}}, t)=\int_V G_l({\bf{r}})  {\bf B}({\bf{x+r}}, t) d^3r,
\end{equation}
where $G_l({\bf{r}}) =l^{-3} G({\bf{r}}/l) $ with $G({\bf{r}})$ being a smooth and rapidly decaying kernel, e.g., the Gaussian kernel $G(r)\sim e^{-r^2}$. We will call $\bf B$ the bare field whereas ${\bf B}_l$ is the renormalized, or coarse-grained, field at scale $l$.\footnote{Without loss of generality, we also assume $G({\bf{r}})\geq 0$, $\lim_{|\bf r|\rightarrow \infty} G({\bf{r}})\rightarrow 0$, $\int_V d^3r G({\bf{r}})=1$, $\int_V d^3r \; {\bf{r}}\;G({\bf{r}})=0$, $\int_V d^3r |{\bf{r}}|^2 \;G({\bf{r}})= 1$ and $G({\bf{r}})=G(r)$ with $|{\bf{r}}|=r$. Mathematically, $G \in C_c^\infty ({\mathbb{R}})$; the space of infinitely-differentiable functions with compact support. A function $g$ is said to have a compact support (set of its arguments for which $g\neq 0$) if $g=0$ outside of a compact set (equivalent to closed and bounded sets in ${\mathbb{R}}^m$). } Integration by parts in $\nabla_x{\bf B}_l$ makes the derivative act on $G$, implying that ${\bf B}_l$ is Lipschitz-continuous even if ${\bf B}$ is not. Differential equations containing the field $\bf B$ can also be multiplied by the kernel $G$ and integrated to get the corresponding renormalized equations. The renormalized induction equation thus reads

\begin{equation}\label{induction1}
{\partial {\bf{B}}_l\over \partial t}=\nabla\times ( {\bf{u}}_l \times {\bf{B}}_l - {\bf{R}}_l-{\bf P}_l),
\end{equation}

using the renormalized Ohm's law ${\bf E}_l+({\bf u\times B})_l={\bf P}_l$, which can also be written as 

\begin{equation}\label{Ohm}
{\bf{E}}_l={\bf P}_l+{\bf R}_l-{\bf{ u}}_l \times {\bf{B}}_l.
\end{equation}
 Even with a negligible non-ideal term $\bf P$, the non-linear term ${\bf{R}}_l=-( {\bf{ u \times B}})_l+{\bf{u}}_l \times {\bf{B}}_l$ can be large. This term, known as the turbulent electromotive force (EMF) ${\cal{E}}_l\equiv-{\bf{R}}_l$, is the motional electric field induced by turbulent eddies of scales smaller than $l$ and plays a crucial role in magnetic dynamo theories. One can use the renormalized induction equation, eq.(\ref{induction1}), to study the time evolution of the unit tangent vector, $\hat{\bf B}_l={ {\bf B}_l / B_l }$ (related to magnetic topology) and magnitude $B_l=|{\bf B}_l|$ (related to magnetic energy) separately \cite{JV2019}. The induction equation implies

  \begin{equation}\label{zap10}
  \begin{cases}
  \partial_t \hat {\bf{B}}_l={\nabla\times({\bf u}_l\times{\bf B}_l)_\perp \over B_l}-({{\bf{\Sigma}}_l}_\perp+{{\boldsymbol{\sigma}}_l}_\perp),\\
    \partial_t B_l=\nabla\times({\bf u}_l\times{\bf B}_l)_\parallel-B_l({{\bf{\Sigma}}_l}_\parallel+{{\boldsymbol{\sigma}}_l}_\parallel),
  \end{cases}
 \end{equation}
where $(.)_\parallel$ and $(.)_\perp$ respectively refer to the parallel and perpendicular direction with respect to ${\bf B}_l$ and 
\begin{equation}\label{slipV}
\begin{cases}
{\bf{\Sigma}}_l={(\nabla\times{\bf{R}}_l) \over B_l},\\
{\boldsymbol{\sigma}}_l={(\nabla\times{\bf{P}}_l) \over B_l}.
\end{cases}
\end{equation}
These terms, as we will see in the next section, are associated with field-fluid slippage and reconnection. It is also important to note that ${\boldsymbol{\sigma}}_\perp, {\bf{\Sigma}}_\perp$ at any scale $l$ are associated with non-ideal plasma and non-linear turbulent phenomena affecting the field's topology whereas ${\boldsymbol{\sigma}}_\parallel, {\bf{\Sigma}}_\parallel$ dissipate magnetic energy without any direct effect on the evolution of magnetic topology.

Let us also shortly comment on the concept of field lines or integral curves of a given magnetic field $\bf B$. At time $t_0$, the solution of the following initial value problem is a curve,  a magnetic filed line, passing through an arbitrary point $\bx$:

 \begin{equation}\label{fieldline4}
\begin{cases}
{\partial\bxi_{\bx}(s, t_0)\over \partial s} =\hat{\bf B}(\bxi_\bx(s, t_0), t_0),\\
 \bxi_\bx(0, t_0)=\bx. 
 \end{cases}
\end{equation} 
This initial value problem has a unique solution\footnote{This is the Picard-Lindel{\"o}f theorem for a system of differential equations.} if $\hat{\bf B}({\bx}, t)$ is uniformly Lipschitz in $\bx$ \cite{dynamicsJV2019}, which may not hold in turbulence (see below). Moreover, even with a unique solution at time $t_0$, the field lines will not necessarily continuously evolve in time unless $\hat{\bf B}$ is uniformly Lipschitz in spacetime position vector $\vec x=({\bx}, t)$ (see Appendix A  and \cite{dynamicsJV2019} for details). The physical implication is that magnetic field lines do not generally deform in a continuous manner in dissipative and turbulent media. It is indeed a commonplace oversimplification to formulate or describe phenomena such as reconnection in such environments appealing to the notion of field lines, which are abstract mathematical, and not physical, objects useful only in certain situations. Nevertheless, to be very clear, it should be emphasized that the H{\"o}lder singularities of magnetic and velocity fields in MHD (\cite{Eyinketal2013}; \cite{Eyink2018}; \cite{JV2019}) are usually asymptotic and the fields are in fact differentiable at very small scales. But, as a general remark, one should keep in mind the simple fact that even the concept of a fluid may lose its meaning if one naively goes down to extremely small scales. On the other hand, and more importantly, even at scales which are large enough to let the fluid approximation be valid but otherwise are very small, e.g., compared to the system size or down in the turbulence inertial range, an extreme sensitivity to initial conditions (usually characterized by Lyapunov exponents) may plague differential equations such as the initial value problem that defines magnetic field lines; eqs.(\ref{fieldline4}). Take for example two field lines, $\bxi_{\bx'}(s)$ and $\bxi_{\bx}(s)$ starting from two nearby points $\bx$ and $\bx'$. With a small but still non-zero magnetic diffusivity, at very small scales, the field can be Lipschitz and its corresponding field lines unique. Nevertheless, these integral curves, in general, will show extreme sensitivity to the initial conditions such that the distance between them, i.e., $|\bxi_{\bx'}(s)-\bxi_{\bx'}(s)| $, may become independent of  $\bx'-\bx$ at distances $s\gg |\bx'-\bx|$. The corresponding large Lyapunov exponents indicate explosive separation of magnetic field lines which will not be present if the field lines exhibit standard deterministic chaos in which, unlike turbulence, the system preserves a memory of the initial conditions (see also \cite{Eyink2011}; \cite{Eyink2015}; \cite{Eyink2018}). Finally, one might argue that magnetic fields lines can be considered well-defined at an instant of time. This is true and probably useful under certain conditions, however, one should keep in mind that in general, even with continuous equations of motion, the field lines defined at two different times $t_1$ and $t_2$ will be completely different objects.

\section{Slippage and Reconnection}\label{section3}

The non-ideal effects in the renormalized Ohm's law, eq.(\ref{Ohm}), denoted collectively by ${\bf P}_l$ at scale $l$, arise from micro-scale plasma effects such as the resistive electric field, Hall effect etc., which drive reconnection at small diffusive scales and are mathematically represented by ${{\boldsymbol{\sigma}}_l}$ in the induction equation; eqs.(\ref{zap10}). On the other hand, the non-linear term ${\bf R}_l$ in the Ohm's law arises from non-linear interactions below the arbitrary scale $l>0$ which correspond to the non-linear term ${{\boldsymbol{\Sigma}}_l}$ in the induction equation; eqs.(\ref{zap10}). At larger scales in the turbulent inertial range, $\Sigma_l$ dominates $\sigma_l$. However, $\Sigma_l$ decreases with decreasing scale and eventually becomes comparable to $\sigma_l$ at the dissipative scale down the inertial range. Below the dissipative scale, $\sigma_l$ dominates $\Sigma_l$. One may argue that the explosive nature of super-linear Richardson diffusion in the inertial range brings distant field lines to small distances set by resistivity where they may reconnect while it also causes explosive separations between initially close field lines. Nevertheless, this argument, in particular its first part, takes the notion of field lines in a very literal sense which we try to avoid here. The notion of close or distant field lines in real space can be avoided altogether employing instead the phase space trajectories in the context of dynamical systems. What happens in real space, during reconnection, can be understood in terms of the super-linear amplification of small scale magnetic disturbances, generated by the non-ideal plasma effects at the dissipative scale, by the turbulence. This is somehow similar to cosmological inflation which might have stretched sub-atomic quantum fluctuations to astrophysical density perturbations which in turn seed large scale cosmological structures.

Physically, ${\bf B}_l({\bx}, t)$ is the weighted-average magnetic field of a fluid parcel of size $l$ at point $\bx$. Since $G({\br}/l)$ is a rapidly decaying function so the integral ${\bf B}_l({\bf{x}}, t)=\int_V G_l({\bf{r}})  {\bf B}({\bf{x+r}}, t) d^3r$ gets smaller and smaller contributions from points at distances $\gg l$ from $\bx$. If we renormalize the field at a larger scale $L>l$, on the other hand, we will get the average magnetic field of a fluid parcel of scale $L$ at point $\bx$, i.e., ${\bf B}_L({\bx}, t)$ which is in general different from ${\bf B}_l({\bx}, t)$ because the weight function $G({\br}/L)$ gets major contributions only from points with distance $\sim L>l$ from $\bx$. In a laminar flow threaded by a smooth magnetic field with a large curvature radius $\gg L$, we expect $\hat{\bf B}_l.\hat{\bf B}_L\simeq 1$. For a stochastic magnetic field in a fully turbulent medium, on the other hand, $-1\leq \hat{\bf B}_l.\hat{\bf B}_L\leq 1$ becomes a rapidly varying stochastic variable which measures the spatial complexity (or stochasticity level) of $\bf B$ at point $\bf x$. Its root-mean-square (rms) value tells us how spatially complex (or stochastic) the field is on average in a given volume $V$. For a stochastic field, it is also a measure of the field's stochasticity. To obtain a non-negative quantity, we can volume average ${1\over 2}|\hat{\bf B}_l({\bx}, t).\hat{\bf B}_L({\bx}, t)-1|$. In fact, applied to the velocity field $\bf u$ in a turbulent flow, e.g., in a fully turbulent flow, the quantity ${1\over 2}|\hat{\bf u}_l({\bx}, t).\hat{\bf u}_L({\bx}, t)-1|$ measures the spatial complexity (or the level of randomness) of the fluid motions at point $({\bx}, t)$. Here ${\bf u}_l$ (${\bf u}_L)$ is the velocity field renormalized at scale $l$ ($L$). The spatial complexity, or stochasticity level, of the magnetic (or velocity) field in an arbitrary spatial volume $V$ is defined as the root-mean-square of this quantity \footnote{A more general definition employs the ${\cal L}_p$ norm 

$$S_p (t) ={1\over 2} \parallel  \hat{\bf B}_l.\hat{\bf B}_L-1 \parallel_p \equiv {1\over 2} \Big[ \int_V \Big|   \hat{\bf B}_l.\hat{\bf B}_L -1\Big|^p\;{d^3x\over V}\Big]^{1/p}$$
for any arbitrary $p\in \mathbb{N}$. We take $p=2$ which corresponds to rms value; $S_2 (t) ={1\over 2} (  \hat{\bf B}_l.\hat{\bf B}_L-1)_{rms}$. See \cite{JV2019} and \cite{SecondJVV2019} for more details.}\cite{JV2019}:

\begin{equation}
S(t)={1\over 2} (\hat{\bf B}_l.\hat{\bf B}_L-1)_{rms}. 
\end{equation}

In the language of topological dynamics in mathematics, a topological entropy can be defined in the magnetic phase space $(\bf x, B)$, corresponding to magnetic dynamical system defined by $\dot{\bx}={\bf B}$ and $\partial_t {\bf B}=-\nabla\times {\bf E}$ as a measure of the complexity of the system \cite{dynamicsJV2019}. Similarly, in real Euclidean space $\mathbb{R}^3$ (or in general on a manifold), the function $S(t)$ is a measure of spatial complexity of the magnetic field. For stochastic (turbulent) fields, spatial complexity can also be takes as a measure of the magnetic stochasticity level (for mathematical details see  \cite{JV2019} and \cite{dynamicsJV2019}).

Conventional magnetic flux freezing (Alfv\'en theorem), which asserts that the magnetic field is perfectly frozen into a fluid with a vanishingly small resistivity, fails in turbulent flows. However, a more general form known as stochastic flux freezing \cite{Eyink2011} holds in general flows which can be interpreted statistically; the field follows the turbulent fluid motions only in an average and stochastic manner (\cite{JV2019}; \cite{SecondJVV2019}). Therefore, turbulence will tend to tangle the field and increase its spatial complexity (stochasticity); $T(t)=\partial_t S(t)>0$.  On the other hand, the field resists the increase in its spatial complexity (by means of magnetic tension force) and at some point slips through the fluid to relax, during which $T(t)<0$. Thus, when magnetic complexity (stochasticity) level reaches a local maximum, i.e., $T(t)=0$ and $\partial_t T(t)<0$, and then starts to decrease, i.e., $T(t)<0$, the rapidly relaxing field will accelerate fluid particles (by means of Lorentz force). This in turn increases kinetic complexity (stochasticity) level $s(t)$:

\begin{equation}
s(t)={1\over 2} (\hat{\bf u}_l.\hat{\bf u}_L-1)_{rms}.
\end{equation}
This picture associates magnetic reconnection with $T(t)= 0$ and $\partial_t T(t)<0$ which is almost simultaneous with $\tau(t)=\partial_t s(t)>0$---magnetic reconnection enhances turbulence thus kinetic stochasticity.

Quite apart from the mathematical subtleties regarding the field lines and field topology, as discussed above, reconnection or field-fluid slippage in a small volume $\sim l^3$ of the fluid is expected to correspond to spontaneous, large changes in the magnetic direction vector, $\hat{\bf B}_l$, measured with respect to the large scale field $\hat{\bf B}_L$ with $L\gg l$. Likewise, a global topology change in a larger volume $\sim L^3$ corresponds to large changes in the magnetic direction vector $\hat{\bf B}_L$ measured with respect to the small scale field $\hat{\bf B}_l$. In other words, we may use the average of $\hat{\bf B}_l.\partial_t \hat{\bf B}_L+\hat{\bf B}_L.\partial_t \hat{\bf B}_l$ as a measure of the field reconnection at point $({\bx}, t)$. The weighted average of this quantity, which is incidentally equal to $\partial_t( \hat{\bf B}_l. \hat{\bf B}_L )$, with the weight function $w({\bx}, t)=( \hat{\bf B}_l.\hat{\bf B}_L-1)$ is in fact proportional to $T(t)=\partial_t S(t)$;

\begin{equation}\label{Tdeform}
T(t)=  {1\over 4 S(t)} \int_V \;(\hat{\bf B}_l.\hat{\bf B}_L-1)\partial_t (\hat{\bf B}_l.\hat{\bf B}_L)\; {d^3x\over V},
\end{equation}
which represents the rate of change of the magnetic field configuration (but not necessarily rate of change in the velocity field, see eq.(\ref{Tdeform2}) below). A similar quantity can be written for the velocity field, $\tau(t)=\partial_t s(t)$; 
\begin{eqnarray}\label{Tdeform1}
\tau(t)&=&  {1\over   (\hat{\bf u}_l.\hat{\bf u}_L-1)_{rms}} \\\notag
&&\times \int_V \;{1\over 2} (\hat{\bf u}_l.\hat{\bf u}_L-1 )\partial_t (\hat{\bf u}_l.\hat{\bf u}_L)\; {d^3x\over V}.\\\notag
\end{eqnarray}

Combining the first equation in (\ref{zap10}) with (\ref{Tdeform}) shows that $T(t)$ depends on weighted averages of ${{\boldsymbol{\sigma}}_l}_\perp$, ${{\boldsymbol{\Sigma}}_l}_\perp$ and their counterparts at scale $L$. It has already been shown \cite{Eyink2015} that these quantities act as source terms in differential equations governing the relative velocity between the field and fluid. Here, we will avoid such literal interpretations in terms of field lines and their relative motion with respect to the fluid, however, one may still think of ${{\boldsymbol{\sigma}}_l}_\perp$, ${{\boldsymbol{\Sigma}}_l}_\perp$ as a measure of the field-fluid slippage ({\cite{JV2019}; \cite{SecondJVV2019}). 

Fig.(\ref{dphiT2}) plots $T(t)$ and $\tau(t)$ in an incompressible, homogeneous MHD numerical simulation stored online (\citep*{JHTDB};\citep*{JHTB1};\citep*{JHTB2}). Turbulence gradually increases an initially small magnetic stochasticity, hence $T=\partial_t S>0$. At some point, the field's resistance against tangling leads to a spontaneous field-fluid slippage (reconnection) which decreases the magnetic stochasticity from its maximum ($T=\partial_t S=0$ and $\partial_t^2 S=\partial_t T<0$) and accelerates fluid elements (points A, B and C in Fig.(\ref{dphiT2})). This in turn increases kinetic stochasticity, i.e., $\tau=\partial_t s>0$, which corresponds to the positive values of $\tau$ in Fig.(\ref{dphiT2}). A more detailed theoretical and numerical approach to reconnection using this formalism can be found in \cite{SecondJVV2019}.

\begin{figure}
 \begin{centering}
\includegraphics[scale=.55]{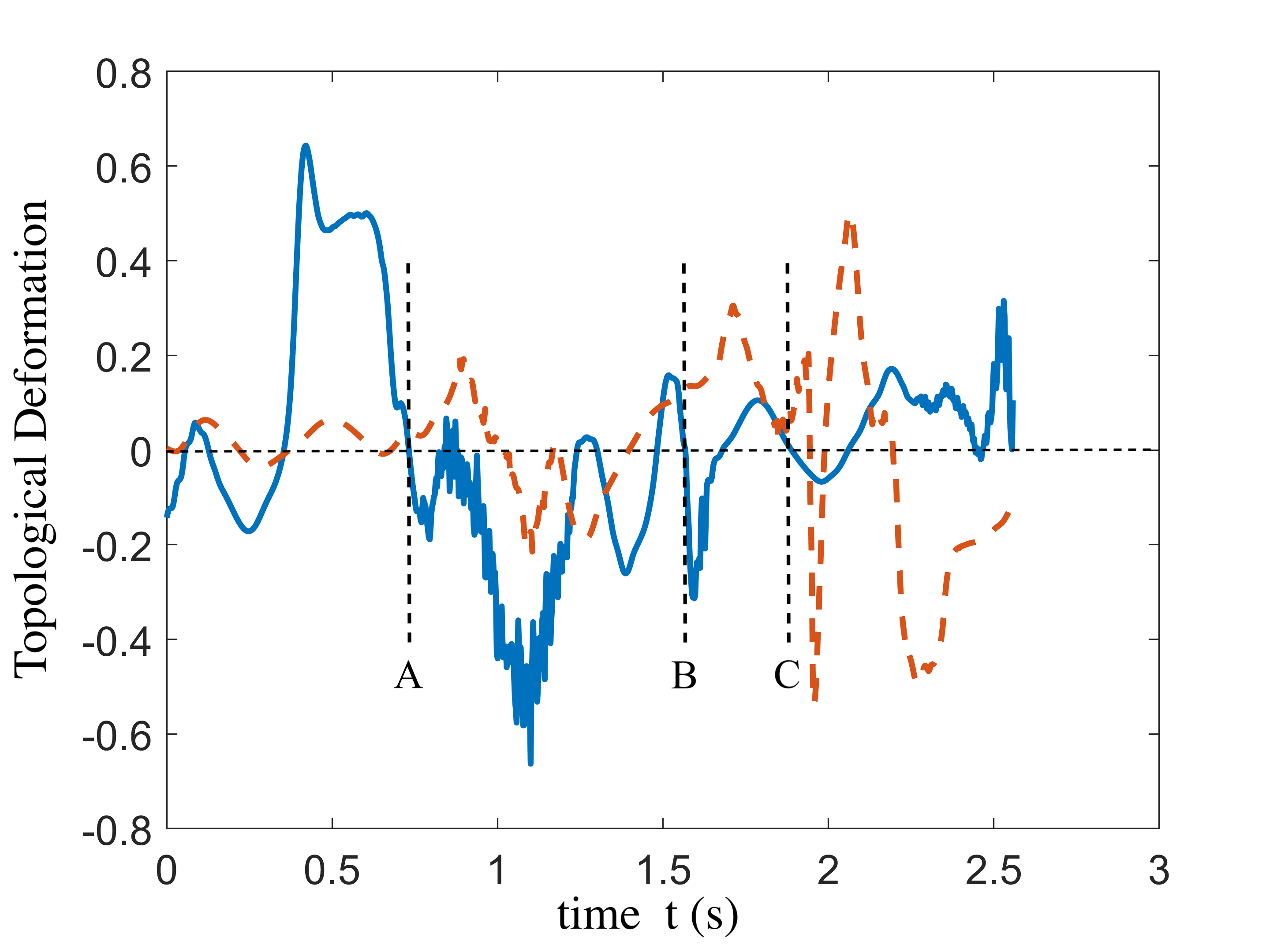}
\caption {\footnotesize {The rate of change of magnetic, $T=\partial_t S$, (blue, solid curve) and kinetic, $\tau=\partial_t s$, (red, dashed curve) spatial complexities in an MHD numerical simulation. As turbulence entangles the magnetic field, the magnetic spatial complexity (stochasticity) $S$ increases hence its time derivative (magnetic topological deformation) becomes positive, $T=\partial_t S>0$. The field's resistance against tangling leads at some point to its slippage through the fluid (reconnection/slippage) which reduces its spatial complexity (stochasticity), $T<0$. Hence, reconnection peaks when $\partial_t S=T=0$ and $\partial_t^2 S=\partial_t T<0$, marked as A, B and C in the graph (local maxima of $S$). This sudden decrease in magnetic spatial complexity  relaxes the field to a smoother configuration (slippage). It may also eject the fluid out of the surrounding region which in turn increases the kinetic stochasticity; $\tau=\partial_t s>0$ (reconnection). Notice the positive values of $\tau$ on the right hand side of A, B and C at which reconnection peaks. }}\label{dphiT2}
\end{centering}
\end{figure}

The statistical formalism presented so far can also be used to define a rate for field-fluid slippage and reconnection. The incompressible Navier-Stokes equation can be written as

\begin{equation}\label{NS1}
{\partial {\bf{u}}\over \partial t}+\nabla.( {\bf u}{\bf u})=\bf f+{\bf j\times B},
\end{equation}
where $\bf j=\nabla\times B$ is the electric current and $\bf f$ represents all other non-magnetic force densities including pressure gradient $\nabla p$, viscous force $\nu\nabla^2\bf u$ and any external force. The renormalized form of this equation reads

\begin{equation}\label{NS3}
{\partial {\bf{u}}_l\over \partial t}+\nabla . ({\bf{u}}_l{\bf{u}}_l +{\bf M}_l)={\bf f}_l+{\bf j}_l\times {\bf B}_l+{\bf N}_l,
\end{equation}
where ${\bf M}_l=({\bf{u}}{\bf{u}}  )_l-{\bf{u}}_l{\bf{u}}_l $ is the turbulent stress tensor. The non-linear term 

\begin{equation}\label{Rfield}
{\bf N}_l=({\bf j\times B})_l-{\bf j}_l\times{\bf B}_l,
\end{equation}
 may be dubbed the reconnection field, for it is the magnetic force responsible for spontaneously driving fluid jets and, consequently, increasing the kinetic stochasticity $s(t)$. In fact, ${\bf N}_l$ is the turbulent magnetic force generated by eddies at scales $<l$. The reconnection rate, at which this force changes the kinetic stochasticity $s(t)$, is determined by the contribution of ${\bf N}_l$ to $\tau=\partial_t s$, that is

\begin{eqnarray}\label{Tdeform2}
\tau(t)\Big|_{rec}&=&{1\over   (\hat{\bf u}_l.\hat{\bf u}_L-1)_{rms}} \\\notag
&&\times\int_V \;{1\over 2} (\hat{\bf u}_l.\hat{\bf u}_L-1 )\Big (\hat{\bf u}_L.{   {{\bf N}_l}_\perp \over u_l}+\hat{\bf u}_l.{  {{\bf N}_L}_\perp \over u_L}  \Big){d^3x\over V},
\end{eqnarray}

where ${(.)_l}_\perp$ and ${(.)_L}_\perp$ denote, respectively, the perpendicular components with respect to ${\bf u}_l$ and ${\bf u}_L$. For a reconnection region of scale $l$, embedded in a system of size $\sim L\gg l$, the global field can be initially assumed undisturbed while the local field ${\bf B}_l$ undergoes reconnection/slippage, in which case the last term in eq.(\ref{Tdeform2}), i.e., $\hat{\bf u}_l.( {{\bf N}_L}_\perp /u_L)$, can be neglected. Hence, the reconnection rate will be

\begin{eqnarray}\label{Tdeform3}
\tau(t)\Big|_{rec}&=& \int_V W({\bx}, t) {   \hat{\bf u}_L.{{\bf N}_l}_\perp \over u_l} {d^3x\over V},
\end{eqnarray}

with the weight function
\begin{equation}
W({\bx}, t)= {1\over 2}{   \hat{\bf u}_l.\hat{\bf u}_L-1 \over   (\hat{\bf u}_l.\hat{\bf u}_L-1)_{rms}  } .
\end{equation}

The formalism presented above is based on the concept of scale split energy density $\psi({\bx}, t)={1\over 2}{\bf B}_l.{\bf B}_L$ which can be written in terms of two other scalar fields in the form $\psi=\chi\phi$. Here $\phi=\hat{\bf B}_l.\hat{\bf B}_L$ is associated with magnetic topology used to define spatial complexity (or stochasticity level) while $\chi={1\over 2}B_l B_L$ is related to magnetic energy and is in fact the geometric mean of the energy densities at scales $l$ and $L$; $\chi=\sqrt{ {B_l^2\over 2} .{B_L^2\over 2}}$. These scalar fields have different evolution equations, obtained using the renormalized induction equation, which can be used to study magnetic field evolution in a statistical context; see \cite{JV2019}. In this picture, field-fluid slippage is defined as spontaneous changes in magnetic topology (appreciable, sudden changes in $S(t)$, magnetic complexity or stochasticity level) while magnetic reconnection involves both topology change and energy dissipation (appreciable changes in $E(t)$, magnetic cross energy) \cite{SecondJVV2019}.

\section{Topology and Dynamical Systems}\label{section4}

The most elementary and most famous example in general topology is perhaps given in terms of the topological equivalence of a coffee mug (which has a handle) and a donut (which has a hole). It goes as follows: one can deform a donut to make a coffee mug and vice versa, without tearing or gluing, while one cannot make a ball out of a donut without tearing or gluing. Thus, a donut is said to be topologically equivalent to a coffee mug but a ball, for example, is a topologically distinct object. A donut can be continuously deformed into a mug preserving all its topological properties, but to make a ball out of the same donut, a non-continuous action which involves tearing or gluing should intervene which will change the donut's topological properties (e.g., the initial hole in the donut, as a topological property, disappears by gluing when making a ball out of it). Mathematically, we can say that the donut can be mapped into the mug in such a way that (i) arbitrarily close points on the donut remain arbitrarily close on the mug, i.e., no tearing or gluing, which demands the map to be continuous, (ii) each point on the mug comes from exactly one point on the donut without leaving out any point, i.e., the map is onto and one-to-one, (iii) the map, or the action of deformation, can be reversed and the mug can also be deformed back into the donut, i.e., the map should be invertible. Such a continuous, onto and one-to-one map with continuous inverse (called a homeomorphism) can continuously map a topological space, e.g., a donut, into another topological space, e.g., a mug, and vice versa, preserving all the topological properties. It goes without saying that there is no homeomorphism mapping a donut into a ball thus these two spaces are not homeomorphic. 

Instead of a donut and a mug, let us take two snapshots of the water surface in a calm pond at times $t_1$ and $t_2$. Are these two surface configurations topologically equivalent and thus can be continuously mapped into each other or something, e.g., a pebble thrown into the pond, might have changed the water surface topology some time between $t_1$ and $t_2$? The water surface can be described using a scalar field, therefore, this thought experiment can be taken as starting point to develop a topology for scalar fields. We are not interested in scalar fields at the moment, instead, let us consider vector fields and think of the changing pattern of the three-dimensional magnetic field in a magnetized plasma: is the field at time $t_1$ topologically equivalent to the field at $t_2$ or it might have been changed some time between $t_1$ and $t_2$, e.g., by magnetic reconnection?

A (metric) topology can be defined for any vector field ${\bf B}({\bf x}, t)$. Suffice to note that such a vector field defines a metric space, i.e., corresponding to any pair of vectors ${\bf B}({\bf x}, t)$ and ${\bf B}({\bf y}, t)$, a distance is defined using the Euclidean vector norm, $||{\bf B}({\bf x}, t)-{\bf B}({\bf y}, t)||\geq 0$. Similar to the points on a donut for which a distance can be defined, in this case vectors at different points have a well-defined distance which allows a metric topology to be established. Note that the distance $||{\bf B}({\bf x}, t)-{\bf B}({\bf y}, t)||$ has nothing to do with distances in real Euclidean space; it is only a measure of how small or large the magnitudes of vectors ${\bf B}({\bf x}, t)$ and ${\bf B}({\bf y}, t)$ are compared to each other and also how they are oriented with respect to each other. The fact that, in the above metric topology, we are missing a measure of how distant the vectors ${\bf B}({\bf x}, t)$ and ${\bf B}({\bf y}, t)$ are in real space (i.e., the value of $|| {\bf x} -{\bf y}||$) is important and at the same time unfavorable for reasons we will discuss presently.

One may take the metric topology defined above and ask whether the time translation, $\hat T: {\bf B}({\bf x}, t)\rightarrow {\bf B}({\bf x}, t+\epsilon)$, which corresponds to the time evolution of the field, maps this topological space at time $t$ to another topologically equivalent space at time $t+\epsilon$. It turns out that even for a well-defined vector field ${\bf B}({\bf x}, t)$, its topology as defined above is not preserved as the field evolves in time, i.e., the time translation $\hat T: {\bf B}({\bf x}, t)\rightarrow {\bf B}({\bf x}, t+\epsilon)$ is not a homeomorphism. In other words, time-dependent vector fields in general change their topology as they evolve in time, and thus the whole picture of invoking vector field topology in order to study problems such as magnetic reconnection in terms of magnetic topology change seems useless. However, as mentioned before, we have missed an important point in the above discussion: the topology defined above is not physically interesting: we have not put any constraint on the positions $\bf x$ and $\bf y$ of these magnetic vectors in real space. Since physically we are interested in magnetic vectors in a given volume of real Euclidean space, e.g., in a fusion device or a part of the solar surface, we need to define the distance between vectors, at a given time $t$, in terms of both $||{\bf B}({\bf x}, t)-{\bf B}({\bf y}, t)||$ and $|| {\bf x} -{\bf y}||$. In other words, the distance (metric) between points in our topological space should be defined e.g., as 

\begin{equation}\label{metric10}
\Delta_{({\bf B(x}, t), {\bf B(y}, t))}=\sqrt{||{\bf B}({\bf x}, t)-{\bf B}({\bf y}, t)||^2+|| {\bf x} -{\bf y}||^2}. 
\end{equation}

This is the metric in the phase space $\bf (x, B)$ \cite{dynamicsJV2019}. In short, the Euclidean vector norm, $||.||$, defines a notion of distance (i.e., metric) for the magnetic field $\bf B$, e.g., $||{\bf B}({\bf x}, t)-{\bf B}({\bf y}, t) ||$ can be taken as the distance between the vectors ${\bf B}({\bf x}, t)$ and ${\bf B}({\bf y}, t)$\footnote{With this metric, the vector field defines a metric space, and since any metric space is a topological space, hence the field is associated with a natural metric topology.}. We are interested, however, in magnetic vectors whose distance is measured not only in vector space but also in real space. That is to say, we are interested in both $|| {\bf B}({\bf x}, t)-{\bf B}({\bf y}, t)  ||$ and $|| {\bx}-{\by}   ||$. For example, in reconnection, we are concerned with the magnetic vectors located in a spatial volume, i.e., the reconnection region. The set of all magnetic vectors ${\bf B}({\bf y}, t)$, at an arbitrary time $t$, whose distance from a given vector ${\bf B}({\bf x}, t)$ is smaller than an arbitrary number $r>0$, satisfy the following condition: \footnote{This is the definition of open balls in a metric topology. In three dimensional Euclidean space, for example, an open ball of radius $r>0$ around point $\bf x$ contains all points $\bf y$ for which $||{\bf x-y}||< r$. Also, note that other forms of distance (metric) can be defined as well, such as $|| {\bf B}({\bf x}, t)-{\bf B}({\bf y}, t)  ||+|| {\bx}-{\by}   ||$, but the exact form of the metric is not important here. Finally, note that a method of nondimensionalization should obviously be applied to ensure that expressions which mix $\bf B$ and $\bf x$ are dimensionally correct.}
\begin{equation}\label{metric11}
\sqrt{ ||{\bx}-{\by}||^2+||{\bf B}({\bf x}, t)-{\bf B}({\bf y}, t)||^2}<r.
\end{equation}

The space of all points $({\bf x}, {\bf B})$ equipped with a measure of distance, like the one given by eq.(\ref{metric10}), is by definition a metric space. It can be easily shown, on the other hand, that any metric space is a topological space. Therefore, in the same fashion that the set of all points on a donut, whose mutual distances are well-defined, define a metric topology, the set of all points $({\bf x}, {\bf B})$ in the phase space $\bf (x, B)$, which have a well-defined measure of distance given by eq.(\ref{metric10}), also define a metric topology. Magnetic field topology at time $t$, therefore, is naturally defined in the phase space $({\bf x}, {\bf B})$. 

\begin{figure}
 \begin{centering}
\includegraphics[scale=.35]{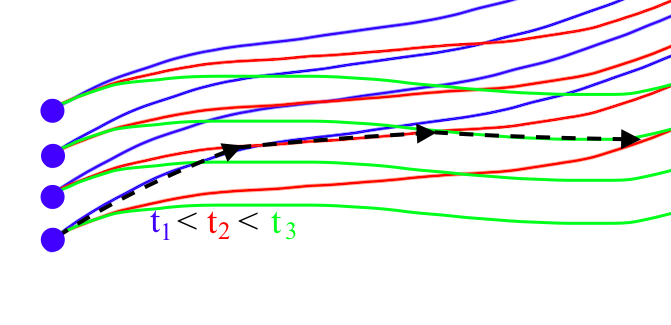}
\caption {\footnotesize {(color online) Three snapshots of magnetic field lines at times $t_3>t_2>t_1$. Each field line is defined as a solution of $d{\boldsymbol \xi}(s)/ds={\hat {\bf B}}({\boldsymbol \xi}(s), t)$ at a given time $t$ with initial condition $\boldsymbol\xi(0)=\boldsymbol \xi_0$ which fixes the field line at some point (solid dots). The motion of a hypothetical particle, or a magnetic vector, with velocity ${\bf B(x}(t), t)$ is also depicted by black, dashed arrows. The particle moves on a path line, i.e., the black, dashed curve ${\bf x}={\bf x}(t)$ with tangent vector $d{\bf x}(t)/dt={\bf B(x}(t), t)$, which completely differs from the field lines. }}\label{pathlines}
\end{centering}
\end{figure}

One may take an arbitrary magnetic field ${\bf B( x}, t)$ as a velocity field and define the trajectory $\bx (t)$ of a hypothetical particle moving with this velocity as
\begin{equation}\label{trajectory1}
{d{\bx}(t)\over dt}={\bf B}({\bx}(t), t);\;{\bx}(0)={\bx}_0,
\end{equation}

where the field solves the induction equation, eq.(\ref{Faraday1}). Eq.(\ref{trajectory1}) defines a trajectory in the phase space $\bf (x, B)$. Also it defines a trajectory in real Euclidean space: it is the trajectory a magnetic vector follows in space as time advances. A snapshot taken from the field ${\bf B}$ at time $t$ will show us all magnetic vectors that define magnetic field lines or streamlines (after fixing each line at a point ${\bf x}_0$ as an initial condition); eq.(\ref{fieldline4}). Alternatively, we can follow magnetic vectors in space as time advances; each vector will move forward on a trajectory defined by eq.(\ref{trajectory1}). In the case of velocity field, these trajectories are called path lines. Thus, it is important to keep in mind the difference between field lines (or streamlines; curves instantaneously tangent to the flow's velocity vector) and path lines (trajectories that individual particles moving with the flow's velocity follow). On the other hand, the above arguments suggest that magnetic reconnection is intimately related to large, positive Lyapunov exponents\footnote{Lyapunov exponents associated with a dynamical system  characterize the rate of separation of initially close trajectories ${\bf x}_1(t)$ and ${\bf x}_2(t)$. Any pair of such trajectories in the phase space, which e.g., solve eq.(\ref{trajectory1}), initially separated by $\delta \mathbf{x}_0$, will diverge at the rate 
$$ | \delta\mathbf{x}(t) | \approx e^{\lambda t} | \delta \mathbf{x}_0 |.$$

with the Lyapunov exponent $\lambda$, largest of which is called the system's maximal Lyapunov exponent (MLE).} corresponding to eq.(\ref{trajectory1}).

\section{The Second Law and Time Symmetry}\label{section5}

With the concept of magnetic topology established in the previous section, we then turn to the question whether or not the topology associated with a given magnetic field is preserved in time. In other words, we may ask if the magnetic topology at time $t$ is equivalent to magnetic topology at a later time $t+\epsilon$. To ensure that time translation keeps the field topology, as it maps the phase space ${\bf B}({\bx}, t)$ into ${\bf B}({\bx}, t+\epsilon)$ with $\epsilon> 0$, it has to be onto, one-to-one and continuous with a continuous inverse, as we discussed before. To have a continuous inverse requires in the first place a well-defined inverse map which takes the field topology backward in time: ${\bf B}({\bx}, t) \rightarrow {\bf B}({\bx}, t-\epsilon)$. To move backward in time requires the equations of motion to be time reversal invariant, i.e. they should not change under time reversal; $t\rightarrow -t$. 

Mathematically, in order to ensure that time evolution preserves the magnetic topology, the field is required to be (i) Lipschitz continuous in $\bx$, (ii) uniformly continuous in $t$, (iii) odd under time reversal ${\bf B}({\bx}, -t)=-{\bf B}({\bx}, t)$ and solve a (iv) time reversal invariant induction equation; see Appendix B or \cite{dynamicsJV2019} for details. 

As for singular fields, e.g., turbulent magnetic fields, one can always renormalize the field and MHD equations to remove any H{\"o}lder singularities, as discussed before. Therefore, the time evolution of the magnetic field, resolved at an arbitrary scale $l>0$ (whose vanishing limit, $l\rightarrow 0$, corresponds to the real, bare field $\bf B$) translates into the dynamics of a particle moving with velocity $\dot{\by}(t)={\bf B}_l({\by}(t), t)$ whose trajectory solves the following equation: 

  \begin{equation}\label{particle1}
  {d^2{\by}\over dt^2}={\bf B}_l.\nabla{\bf B}_l+{\nabla\times({\bf u}_l\times{\bf B}_l)}-B_l({{\boldsymbol{\sigma}}_l}+{{\boldsymbol{\Sigma}}_l}),
 \end{equation}

with appropriate initial conditions. The first term on the right hand side of eq.(\ref{particle1}) is the acceleration due to the magnetic tension force while the second term represents the effects of the flow, and the last two terms are non-ideal plasma and non-linear turbulent effects, respectively. These two last terms break the time symmetry in this equation of motion.

In vacuum, electric field is even under time reversal, ${\bf E}(-t)=+{\bf E}(t)$ while magnetic field is not, i.e., ${\bf B}(-t)=-{\bf B}(t)$. Consequently, the Faraday equation, $\partial_t {\bf B}=-\nabla\times {\bf E}$, respects time reversal invariance as expected. In resistive and turbulent fluids, frequently encountered in astrophysics, however, the time reversal symmetry can be broken by various non-idealities.

Non-ideal terms such as $\bf P$ can break the time symmetry in eq.(\ref{Faraday1}), thus in general ${\bf B}(-t)\neq \pm {\bf B}(t)$. The viscous term similarly breaks the time symmetry in the momentum equation, eq.(\ref{NS1}) implying ${\bf u}(-t)\neq \pm {\bf u}(t)$ and allowing kinetic topology change. This in turn breaks the time symmetry in the induction equation through the term $\bf u\times B$. Consequently, magnetic (as well as kinetic) topology is not preserved in a resistive or viscous fluid. The presence of the motional electric field $\bf u\times B$ in the induction equation and the Lorentz force $\bf j\times B$ in the momentum equation imply that magnetic topology change leads to kinetic topology change and vice versa. A sudden change in magnetic topology can in general accelerate charged particles converting magnetic energy. Hence, even with smooth magnetic and velocity fields, the presence of a non-ideality such as resistivity will prevent magnetic field from keeping its topological properties. It is important to note that turbulent flows are not invariant under time reversal, hence the non-linear term ${\bf R}_l$ also breaks the time symmetry in the induction equation. Super-linear Richardson diffusion in turbulent cascades will in fact amplify any small scale variation in $\hat{\bf B}={\bf B}/|\bf B|$ caused by plasma effects like resistivity.

Time symmetry implies entropy conservation while the second law of thermodynamics indicates that entropy never decreases in isolated systems as time flows forward. Dissipative phenomena, which break time symmetry, are originated in the second law of thermodynamics. Therefore the time symmetry breaking effect of the non-ideal dissipative terms, denoted by $\bf P$ in the Ohm's law, is directly related to the second law. The most fundamental physical law behind magnetic topology change, or reconnection, is the second law of thermodynamics.

\section{Summary and Discussion}\label{section6}
Differential rotation, shear and thermal convection, among other things, produce complex patterns of turbulent flows in magnetized astrophysical systems such as galaxies, stars and accretions disks. Similar situations are encountered in laboratory plasmas too. Such complex flows will in general entangle the field, by means of stochastic flux freezing, in an extremely complicated way. Once very entangled, the built-up magnetic tension will make the field slip through the fluid to reduce its spatial complexity level, otherwise the observed large scale fields in astrophysical objects could never be generated and evolved over cosmological time scales. Serving as a more direct evidence are highly entangled solar magnetic fields which, after escalating the solar activity, are succeeded by smoother magnetic configurations. Such a spontaneous slippage of magnetic fields launching jets of fluid---magnetic reconnection---is usually interpreted and described as a change in the topology of the stochastic magnetic fields in the literature of plasma physics and astrophysics. However, neither magnetic topology nor magnetic stochasticity level is given a precise mathematical definition, and such technical terms are usually used rather loosely. Most often, magnetic field lines are presumed to evolve smoothly in time while magnetohydrodynamic (MHD) equations are assumed to be well-defined, yet without rigorous mathematical justifications. 

In this paper, we have advanced physical arguments to support the idea that the dynamics of magnetic fields in dissipative or turbulent systems can be better considered in terms of magnetic path lines, or the phase space trajectories, rather than magnetic field lines. In fact, as it turns out, for time dependent magnetic fields in dissipative and turbulent fluids, the corresponding field lines do not in general evolve continuously in real space, i.e., the field lines at a given time $t_0$ do not continuously deform to the field lines at a later time $t_1$. This strongly restricts their usefulness in describing, let alone formulating, phenomena such as magnetic reconnection. Magnetic field lines are uniquely defined only if the tangent vector $\hat{\bf B}$ is uniformly Lipschitz continuous, a mathematical condition which is seldom satisfied in astrophysical plasmas. Moreover, even if defined uniquely, field lines are not generally continuous in time unless $\hat{\bf B}$ is uniformly Lipschitz continuous in spacetime vector $\vec x=({\bx}, t)$. Hence appealing to the magnetic field lines as continuously deforming curves in space is mathematically problematic. Trajectories defined by $\dot{\bx}(t)={\bf B}({\bx}(t), t)$, either  in real Euclidean space or in the 6-dimensional phase space $({\bx}, {\bf B})$ with $\bf B$ solving the induction equation, on the other hand, are more fundamental objects which allow one to invoke a variety of standard methodologies in the mathematical theory of dynamical systems, such as stochasticity and entropy, to study magnetic phenomena. 

We have also argued that the term magnetic topology, in the context of reconnection and dynamo theories, is often employed without paying attention to its real mathematical meaning. In principle, topology for a given vector field $\bf B$ can be defined e.g., using the Euclidean norm $||\bf B(x)-B(y)||$ \cite{dynamicsJV2019}. However, even if defined rigorously, e.g., using the fact that a vector field defines a metric space and thus also a topological space, such a topology is not associated with magnetic vectors that are nearby in both vector space and real Euclidean space. The phase space topology, however, satisfies these conditions constituting a more desirable picture in dealing with problems such as magnetic field generation and reconnection which occur in a finite volume of real space. Not only are magnetic topology and topology change well-defined, in terms of physical plausibility, only in the phase space $(\bf x, B)$ but also this formalism leads to a deeper physical understanding of magnetic phenomena in magnetized fluids. These considerations, all in all, establish a vector field topology in a phase space and also suggest the study of magnetic fields in the context of dynamical systems. This approach avoids the difficulties associated with the definition of continuously evolving field lines and leads to a mathematically accurate picture in terms of magnetic topology. In addition, this formalism makes direct contact with fundamental concepts such as entropy and time symmetry in theoretical physics. For instance, the magnetic topology change, which is intimately related to field-fluid slippage and reconnection, turns out to be a consequence of the second law of thermodynamics.

The most important results of this paper can be berieved as follows:

1) The magnetic topology, corresponding to an arbitrary magnetic field ${\bf B(x}, t)$, can be defined as the metric topology of the phase space $(\bf x, B)$. Only with this definition, can magnetic topology be intuitively interpreted as a topological object formed by evolving magnetic vectors in space such that their sudden divergence during reconnection corresponds to a topology change. As the field ${\bf B(x}, t)$ evolves, time translation maps the phase space at time $t_0$ to the phase space at time $t_0+\epsilon$. The phase space topology remains unchanged if this map is a homeomorphism (i.e., a continuous, onto, one-to-one map with continuous inverse). Magnetic topology can change if this map fails to be a homeomorphism.

2) Magnetic topology change is rooted in the fact that the induction and momentum equations in real fluids are not time reversal invariant due to the presence of non-idealities such as resistivity. Magnetic topology change in fluids can manifest itself as field-fluid slippage and reconnection.

3) Reconnection, caused by non-ideal plasma effects such as viscosity and resistivity at a fundamental level, can be extremely amplified by the non-linear effects of turbulence, if present (Richardson diffusion). Although reconnection occurs in both laminar and turbulent flows, but it typically proceeds much faster in turbulence.

3) Reconnection involves spontaneous divergence of magnetic vectors, or path lines ${\bf x}(t)$ with tangent vector $\dot{\bf x}(t)={\bf B(x}(t), t)$, as a result of topology change. This phenomenon can be statistically formulated in terms of the time evolution of magnetic stochasticity (spatial complexity) $S(t)={1\over 2} (\hat{\bf B}_l.\hat{\bf B}_L-1)_{rms}$ and kinetic stochasticity (spatial complexity) $s(t)={1\over 2} (\hat{\bf u}_l.\hat{\bf u}_L-1)_{rms}$. Reconnection is intimately related to the Lyapunov exponents corresponding to $\dot{\bf x}(t)={\bf B(x}(t), t)$ where $\bf B$ solves the induction equation.

\appendix
\section{Continuity of Field Lines}

In this appendix, we obtain the mathematical conditions for the uniform continuity of magnetic field lines in time (for details see \cite{dynamicsJV2019}). The equation defining the integral curves, i.e., magnetic field lines, at time $t_0+\epsilon$ for a real $\epsilon$ reads 
 \begin{equation} \notag
\begin{cases}
{\partial\bxi_{\bx}(s, t_0+\epsilon)\over \partial s} =\hat{\bf B}(\bxi_\bx(s, t_0+\epsilon), t_0+\epsilon),\\
 \bxi_\bx(0, t_0+\epsilon)=\bx. 
 \end{cases}
\end{equation} 

The condition for $\bxi_\bx$ to be continuous in $t$ is $\lim_{\epsilon\rightarrow 0}||\bxi_\bx(s, t_0+\epsilon)-\bxi_\bx(s, t_0)||\rightarrow 0$ for any $s$. We write

\begin{eqnarray}\notag
&& \Big|\Big| \bxi_{\bx}(s, t_0+\epsilon)-\bxi_{\bx}(s, t_0)   \Big|\Big|\\\notag
&&= \Big|\Big|\int_0^s ds'\Big[  \hat{\bf B}(\bxi_\bx(s', t_0+\epsilon), t_0+\epsilon)-\hat{\bf B}(\bxi_\bx(s', t_0), t_0)  \Big] \Big|\Big|\\\notag
&&\leq \int_0^s ds' \Big|\Big|  \hat{\bf B}(\bxi_\bx(s', t_0+\epsilon), t_0+\epsilon)-\hat{\bf B}(\bxi_\bx(s', t_0), t_0)   \Big|\Big|.\\\notag
\end{eqnarray}

Assuming that $\hat{\bf B}$ is Lipschitz in spacetime\footnote{Note that one may also use the Minkowski metric here, which is
$$||\vec{x}_2-\vec{x}_2||=\sqrt{ ||{\bf x}_2-{\bf x}_1||^2 -|t_2-t_1|^2      }.$$
In any case, the continuity of the integral curves in time requires continuity of $\bf B$ in spacetime and not just space.} position vector $\vec{x}=({\bx}, t)$, i.e., 

\begin{eqnarray}\notag
||\hat{\bf B}(\vec{x}_2)-\hat{\bf B}(\vec{x}_1)||&\leq& K_0 ||\vec{x}_2-\vec{x}_2||\\\notag
&=&K_0\sqrt{ ||{\bf x}_2-{\bf x}_1||^2 +|t_2-t_1|^2      },
\end{eqnarray}

 for some $K_0>0$, we can write

\begin{eqnarray}\notag
&& \Big|\Big| \bxi_{\bx}(s, t_0+\epsilon)-\bxi_{\bx}(s, t_0)   \Big|\Big|\\\notag
&&\leq K_0 \int_0^s ds'\sqrt{   \Big|\Big| \bxi_{\bx}(s', t_0+\epsilon)-\bxi_{\bx}(s', t_0)   \Big|\Big|^2+\epsilon^2      }.
\end{eqnarray}
Therefore, we find

\begin{eqnarray}\label{continuity1}
&& {\partial\over\partial s}\Big|\Big| \bxi_{\bx}(s, t_0+\epsilon)-\bxi_{\bx}(s, t_0)   \Big|\Big|\\\notag
&&\leq K_0 \sqrt{   \Big|\Big| \bxi_{\bx}(s, t_0+\epsilon)-\bxi_{\bx}(s, t_0)   \Big|\Big|^2+\epsilon^2      }\\\notag
&&\leq K_0\Big(\Big|\Big| \bxi_{\bx}(s, t_0+\epsilon)-\bxi_{\bx}(s, t_0)   \Big|\Big|+|\epsilon|\Big),
\end{eqnarray}
which implies 
\begin{equation}\label{continuity}
\Big|\Big| \bxi_{\bx}(s, t_0+\epsilon)-\bxi_{\bx}(s, t_0)   \Big|\Big|\leq |\epsilon| \Big( {e^{K_0 s}-1\over K_0} \Big).
\end{equation}
For any finite but arbitrarily large $s>0$, we can take $|\epsilon|$ small enough to make the RHS of (\ref{continuity}) arbitrarily small, which indicates that ${\bxi}_\bx$ is uniformly continuous in time. Consequently, $\bxi_\bx(s, t)$ is uniformly continuous in $t$, provided that $\hat{\bf B}({\bx}, t)$ is Lipschitz in $\vec{x}=({\bx}, t)$. Lipschitz continuity of $\hat{\bf B}$ in $\vec{x}=({\bx}, t)$ indicates that $\hat{\bf B}$ is Lipschitz in both $\bx$ and $t$ which can be seen from the last line of (\ref{continuity1}).

\section{Magnetic Topology}

Here, we obtain conditions under which the magnetic field $\bf B$ will keep its phase space topology for all times \cite{dynamicsJV2019}. The trajectories are solutions of the following non-autonomous differential equation:
 
 \begin{equation}\label{DSeqs2}
\begin{cases}
{d{\bx}(t)\over dt}={\bf B}({\bx}(t), t),\\
{\bx}(t_0)={\bx}_0,
\end{cases}
\end{equation}
 
which has a unique solution if $\bf B$ is uniformly Lipschitz continuous in $\bx$ and continuous in $t$. The time translation operator, acting at any point $({\bx}, {\bf B}({\bx}, t))$ in the phase space, can be represented as

\begin{equation}
\hat{\cal T}_e(\epsilon) ({\bx}, {\bf B}({\bx}, t))=({\bx}, {\bf B}({\bx}, t+\epsilon)).
\end{equation}

It is easy to see that $\hat{\cal T}_e(\epsilon)$ is an onto, one-to-one, and continuous map with continuous inverse. For its continuity, for example, we note that $\hat{\cal{T}}_e(\epsilon)$, for any $\epsilon \in \mathbb{R}$, is continuous (so is its inverse for $\hat{\cal{T}}_e^{-1}(\epsilon)=\hat{\cal{T}}_e(-\epsilon)$) if it is continuous at $\epsilon=0$. In order to show this for any $t$, the following ${\cal L}_1$-norm should vanish in the limit $\epsilon\rightarrow 0$,

\begin{equation}\notag
\lim_{\epsilon\rightarrow 0}  \int_{t\in {\mathbb{I}}_t} dt \Big|\Big| \hat{\cal{T}}_e(\epsilon)({\bx}, {\bf B}({\bx}, t))-\hat{\cal{T}}_e(0)({\bx}, {\bf B}({\bx}, t) )\Big|\Big|.
\end{equation}
Thus the condition for the continuity of $\hat{\cal{T}}_e^{-1}(\epsilon)$ is
\begin{equation}
\lim_{\epsilon\rightarrow 0}  \int_{t\in {\mathbb{I}}_t}  dt \Big|\Big| {\bf B}({\bx}, t+\epsilon)-{\bf B}({\bx}, t) \Big|\Big|\rightarrow 0,
\end{equation}

which follows if $\bf B$ is uniformly continuous in $t$. In order to keep the phase space topology preserved in time, we need to ensure that the phase space at any given time $t_0$, as a topological space,, is homeomorphic to the phase space at another time $t_1$. The condition of continuity for $\hat{\cal{T}}_e^{-1}(\epsilon)=\hat{\cal{T}}_e(-\epsilon)$, on the other hand, requires equations 
\begin{equation}\notag
\begin{cases}
{d{\bx}\over dt}={\bf B}\\
{\partial {\bf B}\over \partial t}=-\nabla\times{\bf E},
\end{cases}
\end{equation}

 to be time reversal invariant, which requires $\bf B$ to be odd, i.e., ${\bf B}({\bx}, -t)=-{\bf B}({\bx}, t)$ and $\bf E$ to be even, i.e., ${\bf E}({\bx}, -t)=+{\bf E}({\bx}, t)$. In  resistive MHD, the Ohm's law contains a non-ideal term $\bf P$, $\bf E=-u\times B+P$, which breaks the time symmetry.

\bibliographystyle{apsrev4-2}
\bibliography{BTopology}

\begin{thebibliography}{19}%
\makeatletter
\providecommand \@ifxundefined [1]{%
 \@ifx{#1\undefined}
}%
\providecommand \@ifnum [1]{%
 \ifnum #1\expandafter \@firstoftwo
 \else \expandafter \@secondoftwo
 \fi
}%
\providecommand \@ifx [1]{%
 \ifx #1\expandafter \@firstoftwo
 \else \expandafter \@secondoftwo
 \fi
}%
\providecommand \natexlab [1]{#1}%
\providecommand \enquote  [1]{``#1''}%
\providecommand \bibnamefont  [1]{#1}%
\providecommand \bibfnamefont [1]{#1}%
\providecommand \citenamefont [1]{#1}%
\providecommand \href@noop [0]{\@secondoftwo}%
\providecommand \href [0]{\begingroup \@sanitize@url \@href}%
\providecommand \@href[1]{\@@startlink{#1}\@@href}%
\providecommand \@@href[1]{\endgroup#1\@@endlink}%
\providecommand \@sanitize@url [0]{\catcode `\\12\catcode `\$12\catcode
  `\&12\catcode `\#12\catcode `\^12\catcode `\_12\catcode `\%12\relax}%
\providecommand \@@startlink[1]{}%
\providecommand \@@endlink[0]{}%
\providecommand \url  [0]{\begingroup\@sanitize@url \@url }%
\providecommand \@url [1]{\endgroup\@href {#1}{\urlprefix }}%
\providecommand \urlprefix  [0]{URL }%
\providecommand \Eprint [0]{\href }%
\providecommand \doibase [0]{https://doi.org/}%
\providecommand \selectlanguage [0]{\@gobble}%
\providecommand \bibinfo  [0]{\@secondoftwo}%
\providecommand \bibfield  [0]{\@secondoftwo}%
\providecommand \translation [1]{[#1]}%
\providecommand \BibitemOpen [0]{}%
\providecommand \bibitemStop [0]{}%
\providecommand \bibitemNoStop [0]{.\EOS\space}%
\providecommand \EOS [0]{\spacefactor3000\relax}%
\providecommand \BibitemShut  [1]{\csname bibitem#1\endcsname}%
\let\auto@bib@innerbib\@empty
\bibitem [{\citenamefont {Schindler}\ \emph {et~al.}(1988)\citenamefont
  {Schindler}, \citenamefont {Hesse},\ and\ \citenamefont
  {Birn}}]{Schindleretal1988}%
  \BibitemOpen
  \bibfield  {author} {\bibinfo {author} {\bibfnamefont {K.}~\bibnamefont
  {Schindler}}, \bibinfo {author} {\bibfnamefont {M.}~\bibnamefont {Hesse}},\
  and\ \bibinfo {author} {\bibfnamefont {J.}~\bibnamefont {Birn}},\ }\href
  {https://doi.org/10.1029/JA093iA06p05547} {\bibfield  {journal} {\bibinfo
  {journal} {Journal of Geophysical Research: Space Physics}\ }\textbf
  {\bibinfo {volume} {93}},\ \bibinfo {pages} {5547} (\bibinfo {year}
  {1988})}\BibitemShut {NoStop}%
\bibitem [{\citenamefont {Biskamp}(1996)}]{Biskamp1996}%
  \BibitemOpen
  \bibfield  {author} {\bibinfo {author} {\bibfnamefont {D.}~\bibnamefont
  {Biskamp}},\ }\href {https://doi.org/10.1007/BF00645113} {\bibfield
  {journal} {\bibinfo  {journal} {Astrophysics and Space Science}\ }\textbf
  {\bibinfo {volume} {242}},\ \bibinfo {pages} {165} (\bibinfo {year}
  {1996})}\BibitemShut {NoStop}%
\bibitem [{\citenamefont {{Priest}}\ and\ \citenamefont
  {{Forbes}}(2007)}]{Priestetal2007}%
  \BibitemOpen
  \bibfield  {author} {\bibinfo {author} {\bibfnamefont {E.}~\bibnamefont
  {{Priest}}}\ and\ \bibinfo {author} {\bibfnamefont {T.}~\bibnamefont
  {{Forbes}}},\ }\href@noop {} {\emph {\bibinfo {title} {Magnetic Reconnection,
  by Eric Priest , Terry Forbes, Cambridge, UK: Cambridge University Press,
  2007}}}\ (\bibinfo {year} {2007})\BibitemShut {NoStop}%
\bibitem [{\citenamefont {Yamada}\ \emph {et~al.}(2010)\citenamefont {Yamada},
  \citenamefont {Kulsrud},\ and\ \citenamefont {Ji}}]{Yamadaetal2010}%
  \BibitemOpen
  \bibfield  {author} {\bibinfo {author} {\bibfnamefont {M.}~\bibnamefont
  {Yamada}}, \bibinfo {author} {\bibfnamefont {R.}~\bibnamefont {Kulsrud}},\
  and\ \bibinfo {author} {\bibfnamefont {H.}~\bibnamefont {Ji}},\ }\href
  {https://doi.org/10.1103/RevModPhys.82.603} {\bibfield  {journal} {\bibinfo
  {journal} {Rev. Mod. Phys.}\ }\textbf {\bibinfo {volume} {82}},\ \bibinfo
  {pages} {603} (\bibinfo {year} {2010})}\BibitemShut {NoStop}%
\bibitem [{\citenamefont {{Lazarian}}\ and\ \citenamefont
  {{Vishniac}}(1999)}]{LazarianandVishniac1999}%
  \BibitemOpen
  \bibfield  {author} {\bibinfo {author} {\bibfnamefont {A.}~\bibnamefont
  {{Lazarian}}}\ and\ \bibinfo {author} {\bibfnamefont {E.~T.}\ \bibnamefont
  {{Vishniac}}},\ }\href {https://doi.org/10.1086/307233} {\bibfield  {journal}
  {\bibinfo  {journal} {\apj}\ }\textbf {\bibinfo {volume} {517}},\ \bibinfo
  {pages} {700} (\bibinfo {year} {1999})},\ \Eprint
  {https://arxiv.org/abs/astro-ph/9811037} {astro-ph/9811037} \BibitemShut
  {NoStop}%
\bibitem [{\citenamefont {{Eyink}}(2015)}]{Eyink2015}%
  \BibitemOpen
  \bibfield  {author} {\bibinfo {author} {\bibfnamefont {G.~L.}\ \bibnamefont
  {{Eyink}}},\ }\href {https://doi.org/10.1088/0004-637X/807/2/137} {\bibfield
  {journal} {\bibinfo  {journal} {\apj}\ }\textbf {\bibinfo {volume} {807}},\
  \bibinfo {eid} {137} (\bibinfo {year} {2015})},\ \Eprint
  {https://arxiv.org/abs/1412.2254} {arXiv:1412.2254 [astro-ph.SR]}
  \BibitemShut {NoStop}%
\bibitem [{\citenamefont {{Jafari}}\ \emph {et~al.}(2018)\citenamefont
  {{Jafari}}, \citenamefont {{Vishniac}}, \citenamefont {{Kowal}},\ and\
  \citenamefont {{Lazarian}}}]{Jafarietal2018}%
  \BibitemOpen
  \bibfield  {author} {\bibinfo {author} {\bibfnamefont {A.}~\bibnamefont
  {{Jafari}}}, \bibinfo {author} {\bibfnamefont {E.~T.}\ \bibnamefont
  {{Vishniac}}}, \bibinfo {author} {\bibfnamefont {G.}~\bibnamefont
  {{Kowal}}},\ and\ \bibinfo {author} {\bibfnamefont {A.}~\bibnamefont
  {{Lazarian}}},\ }\href {https://doi.org/10.3847/1538-4357/aac517} {\bibfield
  {journal} {\bibinfo  {journal} {\apj}\ }\textbf {\bibinfo {volume} {860}},\
  \bibinfo {eid} {52} (\bibinfo {year} {2018})}\BibitemShut {NoStop}%
\bibitem [{\citenamefont {{Jafari}}\ and\ \citenamefont
  {{Vishniac}}(2018)}]{JafariandVishniac2018}%
  \BibitemOpen
  \bibfield  {author} {\bibinfo {author} {\bibfnamefont {A.}~\bibnamefont
  {{Jafari}}}\ and\ \bibinfo {author} {\bibfnamefont {E.}~\bibnamefont
  {{Vishniac}}},\ }\href@noop {} {\bibfield  {journal} {\bibinfo  {journal}
  {arXiv e-prints}\ } (\bibinfo {year} {2018})},\ \Eprint
  {https://arxiv.org/abs/1805.01347} {arXiv:1805.01347 [astro-ph.HE]}
  \BibitemShut {NoStop}%
\bibitem [{\citenamefont {Jafari}\ and\ \citenamefont
  {Vishniac}(2019{\natexlab{a}})}]{JV2019}%
  \BibitemOpen
  \bibfield  {author} {\bibinfo {author} {\bibfnamefont {A.}~\bibnamefont
  {Jafari}}\ and\ \bibinfo {author} {\bibfnamefont {E.}~\bibnamefont
  {Vishniac}},\ }\href {https://doi.org/10.1103/PhysRevE.100.013201} {\bibfield
   {journal} {\bibinfo  {journal} {Phys. Rev. E}\ }\textbf {\bibinfo {volume}
  {100}},\ \bibinfo {pages} {013201} (\bibinfo {year}
  {2019}{\natexlab{a}})}\BibitemShut {NoStop}%
\bibitem [{\citenamefont {{Eyink}}\ \emph {et~al.}(2013)\citenamefont
  {{Eyink}}, \citenamefont {{Vishniac}}, \citenamefont {{Lalescu}},
  \citenamefont {{Aluie}}, \citenamefont {{Kanov}}, \citenamefont
  {{B{\"u}rger}}, \citenamefont {{Burns}}, \citenamefont {{Meneveau}},\ and\
  \citenamefont {{Szalay}}}]{Eyinketal2013}%
  \BibitemOpen
  \bibfield  {author} {\bibinfo {author} {\bibfnamefont {G.}~\bibnamefont
  {{Eyink}}}, \bibinfo {author} {\bibfnamefont {E.}~\bibnamefont {{Vishniac}}},
  \bibinfo {author} {\bibfnamefont {C.}~\bibnamefont {{Lalescu}}}, \bibinfo
  {author} {\bibfnamefont {H.}~\bibnamefont {{Aluie}}}, \bibinfo {author}
  {\bibfnamefont {K.}~\bibnamefont {{Kanov}}}, \bibinfo {author} {\bibfnamefont
  {K.}~\bibnamefont {{B{\"u}rger}}}, \bibinfo {author} {\bibfnamefont
  {R.}~\bibnamefont {{Burns}}}, \bibinfo {author} {\bibfnamefont
  {C.}~\bibnamefont {{Meneveau}}},\ and\ \bibinfo {author} {\bibfnamefont
  {A.}~\bibnamefont {{Szalay}}},\ }\href {https://doi.org/10.1038/nature12128}
  {\bibfield  {journal} {\bibinfo  {journal} {\nat}\ }\textbf {\bibinfo
  {volume} {497}},\ \bibinfo {pages} {466} (\bibinfo {year}
  {2013})}\BibitemShut {NoStop}%
\bibitem [{\citenamefont {{Lalescu}}\ \emph {et~al.}(2015)\citenamefont
  {{Lalescu}}, \citenamefont {{Shi}}, \citenamefont {{Eyink}}, \citenamefont
  {{Drivas}}, \citenamefont {{Vishniac}},\ and\ \citenamefont
  {{Lazarian}}}]{Lalescuetal.2015}%
  \BibitemOpen
  \bibfield  {author} {\bibinfo {author} {\bibfnamefont {C.~C.}\ \bibnamefont
  {{Lalescu}}}, \bibinfo {author} {\bibfnamefont {Y.-K.}\ \bibnamefont
  {{Shi}}}, \bibinfo {author} {\bibfnamefont {G.~L.}\ \bibnamefont {{Eyink}}},
  \bibinfo {author} {\bibfnamefont {T.~D.}\ \bibnamefont {{Drivas}}}, \bibinfo
  {author} {\bibfnamefont {E.~T.}\ \bibnamefont {{Vishniac}}},\ and\ \bibinfo
  {author} {\bibfnamefont {A.}~\bibnamefont {{Lazarian}}},\ }\href
  {https://doi.org/10.1103/PhysRevLett.115.025001} {\bibfield  {journal}
  {\bibinfo  {journal} {Physical Review Letters}\ }\textbf {\bibinfo {volume}
  {115}},\ \bibinfo {eid} {025001} (\bibinfo {year} {2015})},\ \Eprint
  {https://arxiv.org/abs/1503.00509} {arXiv:1503.00509 [physics.plasm-ph]}
  \BibitemShut {NoStop}%
\bibitem [{\citenamefont {Jafari}\ and\ \citenamefont
  {Vishniac}(2019{\natexlab{b}})}]{dynamicsJV2019}%
  \BibitemOpen
  \bibfield  {author} {\bibinfo {author} {\bibfnamefont {A.}~\bibnamefont
  {Jafari}}\ and\ \bibinfo {author} {\bibfnamefont {E.}~\bibnamefont
  {Vishniac}},\ }\href@noop {} {\bibfield  {journal} {\bibinfo  {journal}
  {arXiv e-prints}\ } (\bibinfo {year} {2019}{\natexlab{b}})},\ \Eprint
  {https://arxiv.org/abs/arXiv:1909.04836} {arXiv:arXiv:1909.04836}
  \BibitemShut {NoStop}%
\bibitem [{\citenamefont {{Jafari}}\ \emph
  {et~al.}(2019{\natexlab{a}})\citenamefont {{Jafari}}, \citenamefont
  {{Vishniac}},\ and\ \citenamefont {{Vaikundaraman}}}]{SecondJVV2019}%
  \BibitemOpen
  \bibfield  {author} {\bibinfo {author} {\bibfnamefont {A.}~\bibnamefont
  {{Jafari}}}, \bibinfo {author} {\bibfnamefont {E.}~\bibnamefont
  {{Vishniac}}},\ and\ \bibinfo {author} {\bibfnamefont {V.}~\bibnamefont
  {{Vaikundaraman}}},\ }\href@noop {} {\bibfield  {journal} {\bibinfo
  {journal} {arXiv e-prints}\ } (\bibinfo {year} {2019}{\natexlab{a}})},\
  \Eprint {https://arxiv.org/abs/1909.04624} {arXiv:1909.04624 [astro-ph.HE]}
  \BibitemShut {NoStop}%
\bibitem [{\citenamefont {{Jafari}}\ \emph
  {et~al.}(2019{\natexlab{b}})\citenamefont {{Jafari}}, \citenamefont
  {{Vishniac}},\ and\ \citenamefont {{Vaikundaraman}}}]{JVV2019}%
  \BibitemOpen
  \bibfield  {author} {\bibinfo {author} {\bibfnamefont {A.}~\bibnamefont
  {{Jafari}}}, \bibinfo {author} {\bibfnamefont {E.}~\bibnamefont
  {{Vishniac}}},\ and\ \bibinfo {author} {\bibfnamefont {V.}~\bibnamefont
  {{Vaikundaraman}}},\ }\href@noop {} {\bibfield  {journal} {\bibinfo
  {journal} {arXiv e-prints}\ } (\bibinfo {year} {2019}{\natexlab{b}})},\
  \Eprint {https://arxiv.org/abs/1908.06474v2} {arXiv:1908.06474v2
  [astro-ph.HE]} \BibitemShut {NoStop}%
\bibitem [{\citenamefont {{Eyink}}(2018)}]{Eyink2018}%
  \BibitemOpen
  \bibfield  {author} {\bibinfo {author} {\bibfnamefont {G.~L.}\ \bibnamefont
  {{Eyink}}},\ }\href@noop {} {\bibfield  {journal} {\bibinfo  {journal} {arXiv
  e-prints}\ } (\bibinfo {year} {2018})},\ \Eprint
  {https://arxiv.org/abs/1803.02223} {arXiv:1803.02223 [physics.flu-dyn]}
  \BibitemShut {NoStop}%
\bibitem [{\citenamefont {{Eyink}}(2011)}]{Eyink2011}%
  \BibitemOpen
  \bibfield  {author} {\bibinfo {author} {\bibfnamefont {G.~L.}\ \bibnamefont
  {{Eyink}}},\ }\href {https://doi.org/10.1103/PhysRevE.83.056405} {\bibfield
  {journal} {\bibinfo  {journal} {\pre}\ }\textbf {\bibinfo {volume} {83}},\
  \bibinfo {eid} {056405} (\bibinfo {year} {2011})},\ \Eprint
  {https://arxiv.org/abs/1008.4959} {arXiv:1008.4959 [physics.plasm-ph]}
  \BibitemShut {NoStop}%
\bibitem [{JHT(2019)}]{JHTDB}%
  \BibitemOpen
  \href@noop {} {\bibinfo {title} {{Forced MHD Turbulence Dataset, Johns
  Hopkins Turbulence Databases}}},\ \bibinfo {howpublished}
  {\url{https://doi.org/10.7281/T1930RBS}} (\bibinfo {year} {2008(Accessed
  April, 2019)})\BibitemShut {NoStop}%
\bibitem [{\citenamefont {{Li}}\ \emph {et~al.}(2008)\citenamefont {{Li}},
  \citenamefont {{Perlman}}, \citenamefont {{Wan}}, \citenamefont {{Yang}},
  \citenamefont {{Meneveau}}, \citenamefont {{Burns}}, \citenamefont {{Chen}},
  \citenamefont {{Szalay}},\ and\ \citenamefont {{Eyink}}}]{JHTB1}%
  \BibitemOpen
  \bibfield  {author} {\bibinfo {author} {\bibfnamefont {Y.}~\bibnamefont
  {{Li}}}, \bibinfo {author} {\bibfnamefont {E.}~\bibnamefont {{Perlman}}},
  \bibinfo {author} {\bibfnamefont {M.}~\bibnamefont {{Wan}}}, \bibinfo
  {author} {\bibfnamefont {Y.}~\bibnamefont {{Yang}}}, \bibinfo {author}
  {\bibfnamefont {C.}~\bibnamefont {{Meneveau}}}, \bibinfo {author}
  {\bibfnamefont {R.}~\bibnamefont {{Burns}}}, \bibinfo {author} {\bibfnamefont
  {S.}~\bibnamefont {{Chen}}}, \bibinfo {author} {\bibfnamefont
  {A.}~\bibnamefont {{Szalay}}},\ and\ \bibinfo {author} {\bibfnamefont
  {G.}~\bibnamefont {{Eyink}}},\ }\href
  {https://doi.org/10.1080/14685240802376389} {\bibfield  {journal} {\bibinfo
  {journal} {Journal of Turbulence}\ }\textbf {\bibinfo {volume} {9}},\
  \bibinfo {eid} {N31} (\bibinfo {year} {2008})},\ \Eprint
  {https://arxiv.org/abs/0804.1703} {arXiv:0804.1703 [physics.flu-dyn]}
  \BibitemShut {NoStop}%
\bibitem [{\citenamefont {Perlman}\ \emph {et~al.}(2007)\citenamefont
  {Perlman}, \citenamefont {Burns}, \citenamefont {Li},\ and\ \citenamefont
  {Meneveau}}]{JHTB2}%
  \BibitemOpen
  \bibfield  {author} {\bibinfo {author} {\bibfnamefont {E.}~\bibnamefont
  {Perlman}}, \bibinfo {author} {\bibfnamefont {R.}~\bibnamefont {Burns}},
  \bibinfo {author} {\bibfnamefont {Y.}~\bibnamefont {Li}},\ and\ \bibinfo
  {author} {\bibfnamefont {C.}~\bibnamefont {Meneveau}},\ }in\ \href@noop {}
  {\emph {\bibinfo {booktitle} {Proceedings of the 2007 ACM/IEEE Conference on
  Supercomputing}}},\ \bibinfo {series and number} {SC '07}\ (\bibinfo
  {publisher} {ACM},\ \bibinfo {year} {2007})\ pp.\ \bibinfo {pages}
  {23:1--23:11}\BibitemShut {NoStop}%
\end{thebibliography}%

\end{document}